\documentclass[preprint,showpacs,showkeys,amsmath,amssymb]{revtex4}
\usepackage{graphicx}% Include figure files

\begin{document}
\title{Age of the Universe, Average Deceleration Parameter and Possible Implications for the End of Cosmology}
\author{J. A. S. Lima}
\email{limajas@astro.iag.usp.br} \affiliation{Universidade de S\~ao
Paulo, Instituto de Astronomia, Geof\'\i sica e Ci\^encias
Atmosf\'ericas \\ Rua do Mat\~ao, 1226, CEP 05508-900, S\~ao Paulo,
SP, Brazil}
\date{\today}
\begin{abstract}
A new expression to the total age of the Universe is derived in terms of the average deceleration parameter. This kinematic result holds regardless of the curvature of the universe as well as of the underlying gravity theory. It remains valid even in the context of brane-world motivated cosmologies.  Since the present age parameter of the Universe is accurately adjusted to  $H_0t_0 = 1$, it is shown that the time averaged value of the deceleration parameter is zero. This also means that the cosmic age today is exactly the one predicted by a relativistic flat cosmological model filled by K-matter, a fluid satisfying the equation of state $p = -{\frac{1}{3}}\rho$. By assuming the validity of this relation (in an average long time meaning), it is argued that the decelerating stages of the expansion must exactly be compensated by the accelerated phases, as if the observed Universe coasts forever. If this is true, the present accelerating stage must be followed by a subsequent decelerating phase as predicted by some recent scalar field and brane-world motivated cosmologies. 
\end{abstract}
 \maketitle

%\PACS{95.36.+x, 98.80.-k}

\bigskip

\section{Introduction}

The  current idea of an accelerated Universe driven by dark energy
is based on a  large convergence of independent
observational results, and constitutes one of the greatest
challenges for our current understanding of fundamental physics
\cite{review}. Among a number of possibilities to describe this
dark energy component, the simplest and most theoretically
appealing way is by means of a cosmological constant $\Lambda$,
which acts on the Einstein field equations as an isotropic and
homogeneous source with a constant equation of state (EoS) parameter, $w
\equiv p/\rho = -1$. The present 
concordance cosmological model (CCM) supported by  all the existing observations  is a flat $\Lambda$CDM model with a matter fraction of $\Omega_{\textrm{matter}}=\Omega_{\textrm{b}}+\Omega_{\textrm{m}} \sim 0.26$ and a vacuum energy contribution of $\Omega_{\Lambda}\sim 0.74$ \cite{spergel}, where $\Omega_{\textrm{b}}$, $\Omega_{\textrm{m}}$ stand for the baryonic and cold dark matter, respectively. 

On the other hand, for any physically relevant model, two important observational quantities are defined by $H_0$ and $q_0$, the present values of the Hubble ($H={\dot R}/R$) and deceleration parameters ($q=-{R\ddot R}/{\dot R}^{2}$), respectively. The first quantity sets the present time scale of the expansion  while the second one is telling us that the present stage is speeding up instead of slowing down as expected before the Supernovae type Ia observations \cite{Perl}. 

In this letter, a new formula to the  total age of the Universe is derived in terms of $H_0$ and the average deceleration parameter, $\bar q$. This expression is a kinematic consequence, and, as such,  it holds regardless of the curvature of the universe, as well as of the underlying gravity theory. As widely known, for  the present composition of the Universe suggested by the CCM, the age of the Universe is accurately described by the simple condition $H_0t_0=1$, thereby yielding for $H_0 = 71$ $Km.s^{-1}/Mpc$, a total age 13.7 Gyrs. As we shall see, this means that after more than 13 Gyrs, the average deceleration parameter is zero.  By assuming that such a fact is not a cosmological coincidence, it is argued that the decelerating stages of the expansion are (in a long time averaging meaning) exactly compensated by the accelerated phases, as if the Universe coasts forever. In particular, this explains why the scale factor seems  to be oscillating  around the coasting solution  driven by K-matter, a fluid satisfying the equation of state $p = -{\frac{1}{3}}\rho$\cite{Kolb} (in this connection see also Refs. \cite{Varun90,Allen, Deepak}). 

\section{Age and the averaging deceleration parameter}

It is now widely believed that the universe has undergone some stages of decelerating ($q >0$) and accelerating ($q<0$) regimes. Since $q$ is a slowly time varying quantity, it is interesting to know what kind of information about the expansion can be inferred by averaging it with respect to time, that is, without the necessity of actually integrating the equation of motion driving the scale factor. In order to see that, let us define the present day average  parameter, $\bar q$, by the following expression

\begin{equation}
{\bar q(t_0)}={\frac{1}{t_0}}{\int^{t_0}_{0}}q(t)dt.
\end{equation} 

By inserting the definition of the deceleration parameter

\begin{equation}
q(t) = - \frac{R\ddot R}{{\dot R}^{2}}= \frac{d}{dt}{\Large[\frac{1}{H}\Large]} - 1,
\end{equation}
it is readily seen that

\begin{equation}
\bar q(t_0) = -1 + \frac{1}{t_0 H_0},
\end{equation}
or still,

\begin{equation}\label{ADA}
t_0 = \frac{H_0^{-1}}{1 + {\bar q}}.
\end{equation}

The meaning of this intriguing expression is manifest.  The present day age of the Universe is proportional to $H_0^{-1}$, as should be expected, but the coefficient depends only on the time average value of the deceleration parameter. It is worth noticing that the above kinematic result holds regardless of the curvature of the universe and its number of fluid components, as well as of the underlying gravity theory (I could not find this simple formula in textbooks or in the literature). The unique condition to be satisfied is that the early universe emerged from a Big-Bang or at least with a very high value of $H_i$.   Now, it is interesting  to check if the above result reproduces the known cases. 

To begin with, let us consider the Einstein Field Equations (EFE) for a general FRW geometry supported  by a one-component fluid:

\begin{equation}\label{FE}
8\pi G\rho = 3{\frac{{{\dot R}}^{2}}{R^{2}}} + 3\frac{K}{R^{2}},
\end{equation}

\begin{equation}\label{PE}
8\pi Gp = -2{\frac{\ddot R}{R}} - {\frac{{\dot R}^{2}}{R^{2}}} - \frac{K}{R^{2}},
\end{equation}
where $R(t)$ is the scale factor and $K=0,\pm 1$ is the curvature parameter. 
In particular, for a flat geometry ($K=0$)
%\begin{equation}\label{mRW}
%ds^2=dt^2 - R^2(t)(dx^2+dy^2+dz^2),
%\end{equation}
and a curvature source satisfying the equation of state, $p=\omega \rho$, the differential equation driving the scale factor reads: 

\begin{equation}
R {\ddot R} + {\large(\frac{1+3\omega}{2}\large)}{\dot R}^{2}= 0,
\end{equation}
whose general solution for $\omega$ constant is:

\begin{equation}\label{SF}
R(t)= R_0 {\large[{\frac{3}{2}}(1 + \omega)H_0t\large]}^{\frac{2}{3(1+\omega)}}.
\end{equation}

The last two equations provide two basic informations. The deceleration parameter and the age of the universe are given by 

\begin{equation}
q = \frac{1+3\omega}{2}=constant\,\,\,\,\,and\,\,\,\,\,  t_0 = \frac{2H_0^{-1}}{3(1+\omega)}.
\end{equation}
On the other hand, since the deceleration parameters are constant for all the above models, their values coincide with the average values and the age can effectively be rewritten as $t_0 = {H_0^{-1}}/({1 + {\bar q}})$ in accordance with (\ref{ADA}). This expression yields for a radiation dominated universe ($\omega = 1/3, \bar q=1$) an age of $H_0 t_0=1/2$ while for dust ($\omega = 0, \bar q=1/2$) we have  $H_0 t_0=2/3$, and, finally, for a flat universe dominated by K-matter ($\omega = -1/3, \bar q=0$) one finds $H_0 t_0=1$. All these results agree with the expression (\ref{ADA}) derived for the average deceleration parameter.  For further reference, we call attention that in the case of a K-matter filled universe, $R(t) = R_0H_0 t$, the age parameter is adjusted to unity at any cosmic time, that is, $Ht\equiv1$. It should also be recalled that another example of coasting model is provided by the Milne empty universe. However, as one may check from equations (\ref{FE}) and (\ref{PE}), the coasting empty relativistic solution can be achieved only if the  geometry of the Universe is a hyperbolic  one ($K=-1$). Given the preference of the WMAP results for a flat Universe with very high confidence level, in what follows we do not consider the coasting Milne cosmological model.

\section {Age of the Universe in $\Lambda$CDM models}

For any relativistic cosmology  based on the Friedmann-Robertson-Walker (FRW) metric, the age of the universe in terms of the redshift $z$ is given by the expression 

\begin{figure}
\begin{center}
\includegraphics[width=85mm]{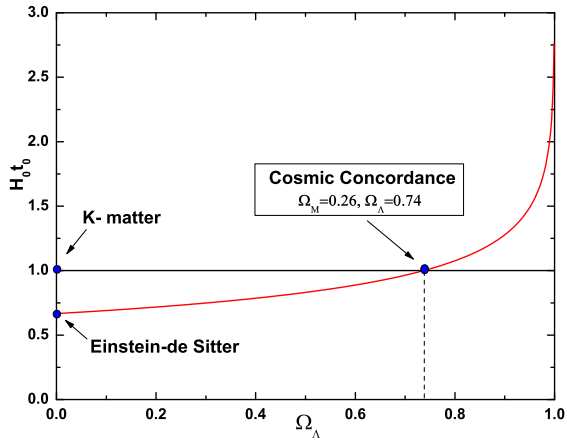}
{\vskip -1.1cm}
\caption{\label{fig1} Age of the Universe in 
$\Lambda$CDM and other cosmologies. The present total age of the Universe for the cosmic concordance model ($\Lambda$CDM) is exactly the same as predicted by the flat coasting K-matter model, $t_0=H_0^{-1}$. Such a coincidence comes from the fact that the $\Lambda$CDM parameters are accurately adjusted to the ones derived from the analysis of the WMAP team.}
\end{center}
\end{figure}

\begin{equation}\label{eq1}
t_0 \equiv {\int_0^{R_0} {\frac{dR}{\dot R}}} = H_0^{-1}\int_0^\infty \frac{dz}{(1+z)E(z)},\,\,\,\,\,\,\,\,\,\,\,\,\,\,\,\,\,\,\,\,\,\,\,\,\,\,\,\,\,\,\,\,\,\,\,\,\,\,\,\,\,\,\,\,\,\,\,\,\,\,\,\,\,\,\,\,\,\,\,\,\,\,\,\,\,\,\,\,\,\,\,\,\,\,\,\,\,\,\,\,\,\,
\end{equation}
\vskip 0.4cm
\noindent where the function $E(z)$ depends on the matter content as well as of the geometry \cite{peebles}. In the general relativistic context, the function $E(z)$ for the flat cosmic concordance model ($\Lambda$CDM)  reads
 
\begin{equation}
E(z)=\left[\Omega_{\textrm{M}}(1+z)^3+ \Omega_{\Lambda} \right]^{1/2},
\end{equation}
where $\Omega_{\textrm{M}}+ \Omega_{\Lambda}=1$. Therefore, the age of the universe depends only on the vacuum density parameter ($\Omega_{{\Lambda}}$) because the matter energy density $\Omega_{\textrm{M}}= 1-\Omega_{{\Lambda}}$ (see Fig. 1). Now, by inserting the values  $\Omega_{\Lambda}=0.74$ and $\Omega_{\textrm{M}}=0.26$, as favored by the three years WMAP collaboration \cite{spergel}, it is easy to show that to the same two digits precision  
\begin{equation}
\int_0^\infty \frac{dz}{(1+z)E(z)}=1,
\end{equation}
thereby showing that the age parameter for the $\Lambda$CDM model is $H_0t_0=1$, which is  exactly the one predicted by the K-matter coasting Universe (in this connection see \cite{MM})). This age is twice the age of a radiation dominated universe, and 51\% larger than the Einstein-de Sitter prediction.  

\begin{figure}
\begin{center}
\includegraphics[width=85mm]{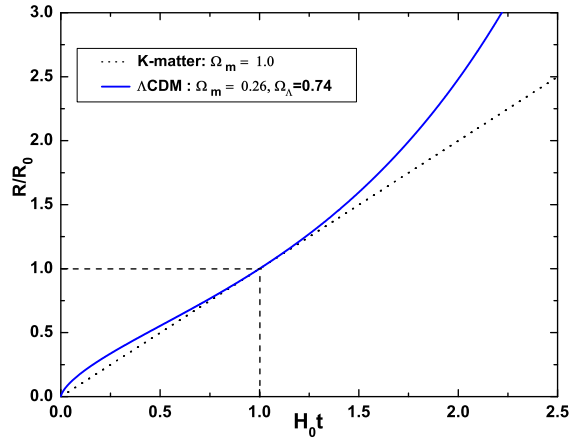}
{\vskip -1.1cm}
\caption{\label{fig2} Evolution of K-Matter and $\Lambda$CDM concordance cosmologies. At present time, $H_0t$=1 and $R(t)/R_0=1$, the scale factor for the K-matter model is tangent to that of the ``standard" $\Lambda$CDM cosmology.}
\end{center}
\end{figure} 

In Fig. 1 we compare the values of the age parameter $H_0t_0$ for a large set of cosmologies, including $\Lambda$CDM, Einstein - de Sitter, and K-matter models.  Effectively, for the ``standard cosmic concordance" model ($\Lambda$CDM), the age of the Universe nowadays is exactly the same one predicted by the coasting K-matter model. This kind of coincidence happens only for the $\Lambda$CDM parameters as derived by the WMAP team analysis.

Such a result can also be seen from a dynamic viewpoint. In Fig. 2 we show the scale factor of the Universe as a function of cosmic time for  K-matter and $\Lambda$CDM models.  As shown there, the two curves are tangent to one another only at the present moment in the whole history of the Universe. In general grounds, one may argue that nowadays the influence of the accelerated expansion of the Universe as described by the $\Lambda$CDM model is exactly compensating the opposite influence of the previous dust dominated  decelerated expansion. However, at light of the results discussed in section (2), it means only that the average deceleration parameter of the ``standard" cosmic concordance $\Lambda$CDM model is zero at the present time. 

Naturally, such a fact may be just an unexpected coincidence. However, the History of Science has already shown that coincidences, mainly in the field of cosmology, deserves  a special attention. Recently, Kutschera and Dyrda \cite{MM} suggested that $H_0t_0=1$ because gravity and antigravity contributions are canceling each other, thereby indicating that the gravitational interaction in the universe has a finite range. Unfortunately, such a correlation is not obvious, and the quoted authors do not provide any further detail for justifying their viewpoint. 

In this letter I will explore a different line of inquire based on the results of section 2. As demonstrated there, this striking result ($H_0t_0=1$) means that the average deceleration parameter determined by taking into account all the curvature sources, $\bar q$, must be identically zero when it is  averaged for a long time interval. Perhaps, due to a still unknown physical reason, the average dynamic behavior of the Universe (after several billions of years) must be just the same of a coasting cosmology. In other words, in an average sense, the scale factor of the Universe coasts forever with the scale factor of the real Universe oscillating slightly around the K-matter solution. Note that equation (\ref{ADA}) can also be rewritten as

\begin{equation}\label{ADA1}
T = \frac{H^{-1}}{1 + {\bar q}},
\end{equation}
where $T$, $H$, and $\bar q$ are, respectively, the age, the Hubble parameter, and the averaging deceleration parameter in a generic cosmic time.  Naturally, $\bar q \neq 0$ for a particular stage of the Universe (dust, radiation, inflation, and so on). Note that $\bar q $ is of the order of unity. Therefore, the above expression shows clearly why the Hubble parameter at any time is the relevant time scale. Further, if the conjecture that $HT \equiv 1$ is valid (after many aeons), it has some interesting consequences that will be discussed next section.   As we shall see, the Universe may evolve trough a cascade of accelerating/deceleration regimes, thereby departing from the present direct extrapolation to a de Sitter final stage predicted by the $\Lambda$CDM model. 

\section {Is the present accelerating phase only a new transient phenomenon?}

It is now widely believed that we live in a remarkable epoch of the cosmic history when the dark energy and matter densities are comparable. Actually, for a Universe one order of magnitude younger it should be impossible to detect any effect of the dark energy on the cosmic expansion. An interesting cosmological question at present, closely related to the ultimate fate of our Universe, concerns the physical effects of a dominant vacuum energy density in the future. If the vacuum is the dark energy component, as assumed in the cosmic concordance model, the Universe will accelerate forever, thereby  evolving inexorably to a de Sitter phase with deceleration parameter $q=-1$. In particular, this means that if one extrapolates the current $\Lambda$CDM to the future, all evidence of the Hubble expansion will disappear with this final state marking the end of cosmology and the return to a static universe. Perhaps still worst, the observers living in the final ``island universe" will be fundamentally incapable of determining the true nature of the Cosmos (for more details see \cite{Krauss} and Refs. therein).

On the other hand, although cosmological scenarios with a $\Lambda$ term might explain most of the current astronomical observations, from the theoretical viewpoint they are plagued with at least two fundamental problems. First, it is really
difficult to reconcile the small value required by observations
($\simeq 10^{-10} \rm{erg/cm^{3}}$) with estimates from quantum
field theories ranging from 50-120 orders of magnitude larger
\cite{weinberg}. Second, eternally accelerating Universes, like the one predicted by a de Sitter model, develops naturally a horizon event in the future, thereby posing challenging questions to the string theory. The basic problem is that the conventional construction of the S-matrix is not  possible because the local observers inside their horizon are not able to isolate particles to be scattered \cite {FLS}. Recently, this conflict inspired some authors to propose a new kind of cosmological scenario driven by a slow rolling homogeneous scalar field \cite{FALR}. The derived equation of state for the field
predicts a transient accelerating phase, in which the Universe was
decelerated in the past, began to accelerate at redshift $z \lesssim 1$, is currently accelerated, but, finally, will return to a deceleration phase in the future. This overall dynamic behavior is profoundly different from the standard $\Lambda$CDM
evolution, and may alliviate some conflicts in reconciling the
idea of a dark energy-dominated universe with observables in
String/M-theory.

\begin{figure}
\begin{center}
\includegraphics[width=105mm]{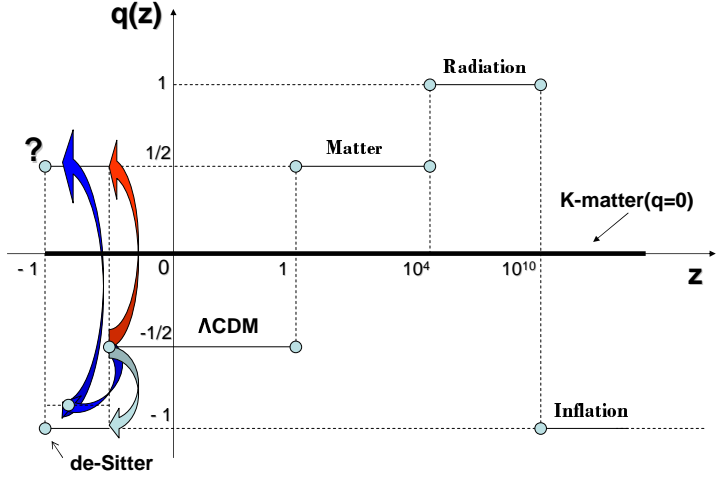}
{\vskip -3.0cm}
\caption{\label{fig3} The deceleration parameter as a function of the redshift (the values of z are not in scale). The Universe emerges from an inflationary stage at the infinite past ($z=\infty$) and evolves to the infinite future ($z=-1$). Abrupt transitions has been assumed for any two subsequent stages. In the actual Universe they are smooth, but, qualitatively, the result is the same, namely, the deceleration parameter of the Universe seems to be oscillating around the K-matter solution ($q=0$). The fate of the Universe is heavily dependent of the next transition in the future. By the arguments presented in the text, the Universe may decelerate (not necessarily mimicking the dust behavior suggested in the diagram).}
\end{center}
\end{figure}

Keeping the above comments in mind, let us now return to the conjecture that $\bar q \equiv 0$ or $HT \equiv 1$ can be true (after many aeons). In other words, that  the Universe behaves (in average) like a K-matter model ($q=0$). Although recognizing the inherent difficulty to probe such a conjecture in the domain of general relativity, I believe that it has some predictive power. In particular, it suggests  that the present accelerating stage does not last forever, and that the possibility to a new transient period for a deceleration regime cannot be neglected.  

In Fig. 3 we sketch this viewpoint in terms of the deceleration parameter in the redshift space. The diagram starts with the inflationary period ($q=-1$), followed by an abrupt transition to the radiation phase ($q=1$). For simplicity, abrupt transitions have also been adopted for any two consecutive phases. Note that in order to have the correct age ($H_0t_0=1$), the average value of the deceleration parameter below $z_t \sim 1$ until the present day must be $-1/2$ (as in the inflationary stage also approximated by a straight line). Now, the problem is to know whether the next transition (at $z<0$) it will be for an accelerating de-Sitter regime ($q=-1$), as required by the $\Lambda$CDM model, or to a decelerating stage which is predicted by some scalar field and brane-world scenarios \cite {FALR,varun03}. As one may check,  the same question appears if we have started with a radiation or matter dominated phase just before the inflationary period. Again, by assuming abrupt transition, the unique alteration in the diagram should be the negative value of the deceleration parameter necessary to have $\bar q=0$ by the present time. In this case, the present transition should be characterized by a deceleration parameter smaller than $-1/2$. 

As one may conclude from Fig. 3, by assuming  $HT\equiv 1$ in the future, there are two possibilities in order to satisfy the constraint $\bar q=0$, both of them pointing to a new deceleration phase, that is,  for positive values of the $q$ parameter. The first is a direct transition for a decelerating regime (red arrow). The second possibility is represented by the two consecutive blue arrows. Initially, the Universes does an intermediate transition to an era with $q$ smaller than $-1/2$, eventually evolving again to a decelerating regime (not necessarily dust dominated as suggested in Fig. 3). For completeness, in Fig. 3 we have also represented the transition to a de Sitter regime ($q=-1$)  as predicted by the $\Lambda$CDM model (green arrow). In this case, it is easy to see that the constraint $\bar q=0$ is not satisfied. 

At this point, we would like to stress that if $\bar q=0$ remains true in the near future, the Universe must evolve to a decelerating regime ($q>0$). Naturally, this does not mean that such a transition should be the last one. Actually, as happened with the transitions in the past, the constraint $HT=1$ is compatible with a sequence of decelerating/accelerating regimes in the future. In principle, one may expect a finite sequence in the redshift range [0,-1] because  from [$\infty, 0$] the Universe seem to be evolved through a small number of distinct stages. In this context, it should be interesting to know if the new transition for a future decelerating regime is somewhat encoded in the Supernovae type Ia data. 
 
\section {Conclusion}

As we have seen, the deceleration parameter of the Universe is a time dependent quantity whose average value  (from $t=0$ to $t=t_0$) is adjusted to zero with a high level of statistical confidence.  This means that the present Universe has the same age of  a coasting cosmology ($t_0=H_0^{-1}$). This happens because nowadays $\bar q =0$. If this is not a coincidence, a new possibility emerges that the observed Universe seems to be oscillating around the trivial FRW type solution, $R(t)=R_0H_0t$, dominated by a flat K-matter model \cite{Kolb,Deepak}. 

The idea of an oscillating Universe is fairly old, and the occurrence of such (pulsating) solutions have fascinated many philosophers and cosmologists since they are associated with the old concept of an eternal return. Actually, since the very beginning of relativistic cosmology such a picture was nicely supported by some exact solutions of the Einstein field equations. Apart from the vacuum case, the cosmic dynamics of  a one-component closed model may be reduced to the one of a global oscillator \cite{lima2}.  However, the Universe is flat and dominated by two components as indicated by the recent observations, and, as such, the idea of an oscillatory Universe must be somewhat rediscussed. 

Based on the general expression derived here, $HT =(1 + \bar q)$, we have advocated that a realistic Universe model should  have a long term oscillation around a coasting flat model driven by K-matter which is dictated by the overall constraint $\bar q=0$. If this is not the case, we are living in a very special time of the $\Lambda$CDM evolution. The physical origin of such a condition is still unknown, but, in principle, it could be determined from a more fundamental cosmological description. As show in Fig. 3, the oscillatory long term behavior becomes more clear when the deceleration parameter is plotted in terms of the redshift. As argued in the last section, the Universe may suffer a new transition to a decelerating phase in the near future, and, probably, such an information is already present in the available observational data (this issue will be discussed in a forthcoming communication).  

\vskip 0.5cm
%\newpage
{\bf Acknowledgments:} The results presented here were obtained during the first Indo-Brazilian Workshop on Cosmology (Pune, India). The nice environment and warm hospitality at IUCCA is acknowledged. I am also grateful to V. Sahni, D. Jain, N. Banerjee, J. S. Alcaniz, I. Waga and R. C. Santos for helpful discussions. This work was partially supported by CNPq and FAPESP, No. 04/13668-0.

\end{document}